\documentclass[aps,pra,singlecolumn,showkeys]{revtex4}
\usepackage{graphicx}
\usepackage{ulem}

\begin{document}

\title{Bistability and all-optical memory in dual-mode diode lasers with time-delayed optical feedback}

\author{N.~Brandonisio, P.~Heinricht, S.~Osborne, A.~Amann, and S.~O'Brien}

\affiliation{Tyndall National Institute, University College Cork, Lee Maltings, Cork, Ireland}  

\begin{abstract}
A proposal for an all-optical memory based on a bistability of single-mode states in a dual-mode diode laser with 
time-delayed optical feedback is presented. The system is modeled using a multimode extension of the Lang-Kobayashi 
equations with injected optical pulses. We uncover the bifurcation structure by deriving analytical expressions for 
the boundaries of the bistable region and demonstrate how the delay time in the external cavity determines an 
optimal pulse duration for efficient switching of the memory element. We also show the relevant role played by 
gain saturation and by the dual-mode solutions of the Lang-Kobayashi equations for the existence of the bistable 
regions. Our results demonstrate that feedback induced bistability can lead to significant performance improvements 
when compared to memory elements based on the injection locking bistability in dual-mode devices.
\end{abstract}


\maketitle

\section{Introduction}

\noindent Applications of all-optical signal processing as an alternative to electronic switching in telecommunication 
and computing networks drives continued interest in implementations of all-optical memory and optical logic 
\cite{Ramos2005}. Many approaches to these problems rely on the co-existence of lasing modes of a given system with 
injected optical pulses for switching between these states. Proposed memories have been based on counter-propagating 
modes in micro-ring devices \cite{Dorren2003, Hill2004}, polarization modes in VCSELs \cite{Mori2006, Mori2008}, 
wavelength bistable states induced by gain cross-saturation \cite{Ishii2005, Raburn2006, Zhukovsky2009} and bistable 
states induced by spatial hole burning effect in DFB lasers \cite{Huybrechts2008}.

Although switching performance in conventional edge-emitting diode lasers is constrained by speed limitations and 
power consumption, these devices represent testbeds for the study of the underlying dynamics based on generic systems 
of equations. Dynamical behaviours found in conventional diode laser systems may be scalable to integrated devices 
based on photonic crystals or microcavities, thus overcoming limitations of speed and power consumption. A recent 
example of this correspondence is provided by a comparison of the structure of the injection locking bistability in 
a dual-mode diode laser and a compact photonic crystal microlaser system that supports two lasing modes 
\cite{Osborne2009, Chen2011}. In both of these systems it was found that optical injection induced a bistability 
between the single mode injection-locked state and a dual-mode equilibrium state of the system. Switching of the 
optical memory was then based on modulated optical injection. Certain weaknesses of this proposal however included 
asymmetric states of the memory and the requirement for the optical injection itself, which played the role of an 
external holding beam. In this paper we propose an alternative but related system, which addresses many of the 
weaknesses of our earlier proposal. We consider wavelength bistability induced in a dual-mode diode laser by 
time-delayed optical feedback. We model this system using an extension of the equations describing the optically 
injected dual-mode system. This proposed memory element has many improved features when compared to devices based 
on injection-locking. These include improved speed, the absence of a holding beam and symmetry of the memory states.

This paper is organized as follows. In section \ref{sec:equations_and_ecms} we study the single-mode and dual-mode
equilibria of the dual-mode system relating these results to the well-known theory of single-mode lasers. In section 
\ref{sec:bistability} we show how bistability between single-mode equilibria appears in the dual-mode system for 
different external cavity lengths. We uncover the bifurcation structure underlying this bistability and derive 
analytical expressions for the boundaries of the bistable region. In section \ref{sec:bistable_regions} we analyze 
the bistable regions varying different parameters of the system and we study the role of the gain saturation in the 
dual-mode system. In section \ref{sec:switching_dynamics} we analyze the switching dynamics of the memory element 
and show how an optimal injected optical pulse duration is determined by the delay time of the system. Our results 
indicate that by scaling the dual-mode device to microlaser dimensions, switching speeds
in the GHz range can be achieved with femtojoule injected pulses.

\section{Dual-mode L-K equations and external cavity modes}
\label{sec:equations_and_ecms}

Our theoretical study is based on a multimode extension of the Lang-Kobayashi equations \cite{Lang1980,Sukow1999, 
Buldu2002}:
\setlength{\arraycolsep}{0.0em}
\begin{eqnarray}
\label{eq:LK_rate_equations}
&& \dot{\tilde{e}}_1=(1/2)(1+i\alpha)(g_1 (2n+1) -1) \tilde{e}_1 
+ \tilde{e}_1^{\mbox{\tiny D}} + \tilde{e}_1^{\mbox{\tiny inj}} \nonumber\\
&& \dot{\tilde{e}}_2=(1/2)(1+i\alpha)(g_2 (2n+1) -1)\tilde{e}_2
+ \tilde{e}_2^{\mbox{\tiny D}} + \tilde{e}_2^{\mbox{\tiny inj}} \\
&& T\dot{n}= p-n-(2n+1)(g_1 |\tilde{e}_1|^2 + g_2 |\tilde{e}_2|^2). \nonumber
\end{eqnarray}
\setlength{\arraycolsep}{5pt}
In the above equations, $\tilde{e}_m$ are the normalized complex electric fields associated with the two modes of 
the laser and $n$ is the normalized excess carrier density in the cavity. 
$\tilde{e}_m^{\mbox{\tiny D}} = \eta\tilde{e}_m(t-\tau)\exp(-i\varphi_m)$ are the normalized complex fields delayed by 
the external cavity round trip time $\tau$. The feedback strength is $\eta$ and $\varphi_m$ is defined such that  
$\varphi_1 = \omega_1 \tau = \varphi$ and $\varphi_2 = \omega_2 \tau = \omega_1 \tau + \Delta\omega \tau = \varphi 
+ \Delta\varphi$, where $\omega_1$ and $\omega_2$ are the free running lasing mode frequencies and  $\Delta\omega = 
\omega_2 - \omega_1$. $\varphi$ and $\Delta\varphi$ are the reference feedback phase and the feedback phase 
difference of the modes. The system is switched using optically injected pulses whose normalized complex fields 
are given by $\tilde{e}_m^{\mbox{\tiny inj}} = k_m\exp(i\Delta\omega_m^{\mbox{\tiny inj}}t)$. Here the detunings 
$\Delta\omega_m^{\mbox{\tiny inj}}$ are measured from the free running lasing mode frequencies. The quantity 
$g_m = g_m^0(1+\epsilon \sum_j \beta_{mj}|\tilde{e}_j|^2)^{-1}$ is the modal nonlinear gain function, where $g_m^0$ 
is the linear modal gain, $\epsilon \beta_{mm}$ is the gain self saturation and $\epsilon \beta_{mj}$ is the gain 
cross saturation. The phase-amplitude coupling is given by $\alpha$, $p$ is the normalized pump current, while 
$T \gg 1$ is the product of the cavity decay rate $\gamma$ with the carrier lifetime $\tau_s$.

Equations (\ref{eq:LK_rate_equations}) have validity within certain ranges of the external cavity length and of 
the feedback strength for two reasons. The first reason is that these equations take into account the propagation 
of the modal fields only in the external cavity and not in the laser cavity. Hence, we assume here that the 
external cavity length needs to be at least one order of magnitude greater than the laser cavity length. The 
second reason is that equations (\ref{eq:LK_rate_equations}) take into account only a single reflection on the 
external mirror. Hence we assume that the feedback strength needs to be less than $\sim 10 \%$, as above this 
value multiple reflections in the external cavity should be taken into account.

For the parameters of the free running device we have used the following values: $g_m^0 = 1$, $\alpha = 2.6$,  
$p = 0.175$ (35 $\%$ above threshold), $\gamma = 0.98 \cdot 10^{12} s^{-1}$, $\tau_s = 0.816 \cdot 10^{-9} s$, 
$T = 800$, $\epsilon = 0.01$, $\beta_{11}=\beta_{22}=1$ and $\beta_{12}=\beta_{21}=\beta=2/3$. These values allow 
for a stable dual-mode equilibrium state in the free running device. The same parameters were successfully 
used to model an optically injected dual-mode system where the primary modes had a frequency spacing of 480 GHz 
and the device length was 350 $\mu$m \cite{Osborne2009, Osborne2009_PRA, OBrien2010}.

Equations (\ref{eq:LK_rate_equations}) allow for two different kinds of solutions: single-mode solutions and 
dual-mode solutions. The single-mode solutions of equations (\ref{eq:LK_rate_equations}) are the external cavity 
modes (ECMs) of the system \cite{Mork1992, Heil2003}. These ECMs are obtained assuming  the ansatz 
$\tilde{e}_m(t) = |\tilde{e}_m(t)| \exp[i (\omega_m + \Delta\omega_m^{\mbox{\tiny sm}}) t]$ in equations 
(\ref{eq:LK_rate_equations}) and zero intensity individually for each modal field. These solutions can be 
regarded as ``equilibrium states'' when considering that their amplitudes and frequencies are constant in time. 
In this paper, we refer to these oscillatory solutions as ``equilibrium states'' in this sense. The single-mode 
ECMs of the dual-mode system are described by the following equations:
\setlength{\arraycolsep}{0.0em}
\begin{eqnarray}
\label{eq:single_mode_equilibrium_feedback}
&& \Delta\omega_m^{\hbox{\tiny sm}}=-\eta\sqrt{1+\alpha^2} \sin(z_m+\arctan(\alpha))\nonumber\\ 
&& I_{m}^{\hbox{\tiny sm}} =
2 ( p + \eta \cos( z_m ) ) / [ ( 2 + \epsilon ) ( 1 - 2 \eta \cos( z_m ) ) ] \\
&& n^{\mbox{\tiny sm}} = ( \epsilon p  - 2 \eta \cos( z_m ) ) / ( 2 + \epsilon ). \nonumber
\end{eqnarray}
\setlength{\arraycolsep}{5pt}
Here $\Delta\omega_m^{\mbox{\tiny sm}}$ is defined as the detuning of each ECM with respect to the frequency of the 
associated free running laser mode, $I_m^{\mbox{\tiny sm}}=|\tilde{e}_m|^2$ is the ECM intensity, and 
$z_m = \Delta\omega_m^{\mbox{\tiny sm}}\tau+\varphi_m$ is the ECM total feedback phase. Each ECM of the system is 
associated with a unique value of the detuning, which can vary from a few gigahertz to tens of gigahertz.

\begin{figure}[t]
\centering
\includegraphics[width=0.9\textwidth]{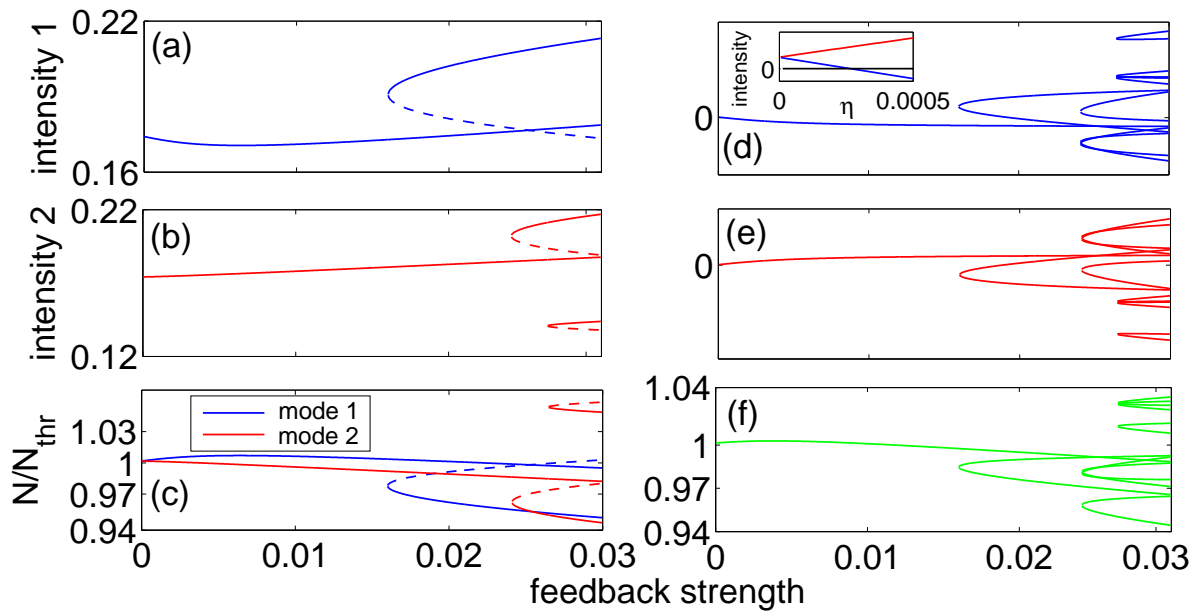}
\caption{(a), (b) Intensities of the single-mode ECMs for a system with 1 cm external cavity length. 
(c) Normalized carrier density of the single-mode ECMs. Continuous and dashed lines indicate  ECMs respectively 
stable and unstable in the single-mode sub-manifolds of the dual-mode system with respect to the saddle-node 
bifurcation. (d), (e) Modal intensities of stable and unstable dual-mode equilibria (the stability is not 
indicated). (f) Normalized carrier density of the dual-mode equilibria. In the inset the modal intensities of 
the dual-mode equilibrium have been plotted for vanishing feedback strength.}
\label{fig:equilibria_1cm}
\end{figure}

The linear stability analysis of the single-mode ECMs of equations (\ref{eq:single_mode_equilibrium_feedback}) 
within the single-mode sub-manifolds of the dual-mode system  demonstrates that the single mode ECMs are born in 
pairs with a stable ECM and an unstable ECM generated through saddle-node bifurcations. This analysis leads to 
the following condition for the stability of the single-mode ECMs \cite{Acket1984, Tromborg1984}:
\setlength{\arraycolsep}{0.0em}
\begin{eqnarray}
\label{eq:saddle_node_condition}
&& \tau\eta\sqrt{1+\alpha^2} \cos(z_m+\arctan(\alpha)) - 1 < 0.
\end{eqnarray}
\setlength{\arraycolsep}{5pt}
When the above condition is satisfied the single-mode ECMs are stable within the single-mode sub-manifolds of 
the dual-mode system with respect to the saddle-node bifurcation, while these ECMs are unstable when this 
condition is greater than zero. Using equations (\ref{eq:single_mode_equilibrium_feedback}) we can study the 
evolution of the single-mode ECMs while varying the feedback strength, as shown in Figure \ref{fig:equilibria_1cm} 
(a), \ref{fig:equilibria_1cm} (b) and \ref{fig:equilibria_1cm} (c). Here we have considered an  external cavity 
length of 1 cm and a laser cavity length of the order of few hundred micrometres. Because of the large primary 
mode spacing, $\varphi$ and $\Delta\varphi$ can be considered free parameters that can be tuned through a relatively 
small displacement of the external mirror. We have chosen the values $\varphi = 1.1 \pi$ and $\Delta\varphi = 0.45 
\pi$ for obtaining the maximum switching speed of the considered device.

In Figure \ref{fig:equilibria_1cm} (a) and \ref{fig:equilibria_1cm} (b) we plot the single-mode ECM intensity, 
while in Figure \ref{fig:equilibria_1cm} (c) we plot the single-mode ECM normalized carrier-density. In these 
figures solid and dashed lines denote respectively stable and unstable ECMs in the single-mode sub-manifolds of 
the dual-mode system. The stability of these ECMs is obtained using equation (\ref{eq:saddle_node_condition}). 
One can also see the presence of saddle-node bifurcations where stable and unstable branches collide.

A fundamental feature of the dual-mode system we consider is the existence of dual-mode solutions, which are 
derived from equations (\ref{eq:LK_rate_equations}), assuming  the ansatz $\tilde{e}_m(t) = |\tilde{e}_m(t)| 
\exp[i (\omega_m + \Delta\omega_m^{\mbox{\tiny sm}}) t]$ and non-zero intensity for both  modal fields. These 
solutions are given by the following equations:
\setlength{\arraycolsep}{0.0em}
\begin{eqnarray}
\label{eq:two_mode_equilibrium_feedback}
&& I_m^{\hbox{\tiny tm}} = [\beta - 1 \pm (x_1^{-1} - \beta x_2^{-1}) 
(2 n^{\mbox{\tiny tm}} + 1)][\epsilon(1-\beta^2)]^{-1} \nonumber\\
&& n^{\mbox{\tiny tm}} = \frac{\epsilon (1-\beta^2) p - (\beta-1)(x_1+x_2) -2 + \beta(x_1 x_2^{-1} 
+ x_2 x_1^{-1})}
{\epsilon (1 - \beta^2) +4 -2 \beta(x_1 x_2^{-1} + x_2 x_1^{-1}) }.
\end{eqnarray}
\setlength{\arraycolsep}{5pt}
Here $x_m = 1 - 2 \eta \cos(\Delta\omega_m^{\mbox{\tiny sm}}\tau+\varphi_m)$, plus and minus signs refer respectively 
to mode 1 and mode 2. dual-mode equilibrium solutions involve two independent detunings which are equal to the 
values determined by the single-mode ECMs. They can therefore be regarded as ECM pairs obtained from nonlinear 
combinations of the single-mode solutions. Because the dual-mode solutions are defined with a single common 
carrier density, they in general involve a redistribution of the intensity from the single-mode values. We have 
used equations (\ref{eq:two_mode_equilibrium_feedback}) for plotting the intensities and the carrier density 
of the dual-mode solutions in Figure \ref{fig:equilibria_1cm} (d), \ref{fig:equilibria_1cm} (e) and 
\ref{fig:equilibria_1cm} (f) as a function of feedback strength with 
the same parameters as in Figure \ref{fig:equilibria_1cm} (a) - (c). In Figure \ref{fig:equilibria_1cm} there 
are regions where the dual-mode solutions are unphysical with negative value of one of the modal intensities. 
One can also see that for each saddle-node of the single-mode ECMs a corresponding saddle-node of dual-mode 
solutions occurs. Furthermore for vanishing feedback strength only a single dual-mode solution exists, which 
corresponds to the stable free running dual-mode equilibrium perturbed by the feedback. This can be seen in 
the inset of Figure \ref{fig:equilibria_1cm}, where the intensities of the two modes converge towards the free 
running dual-mode  equilibrium. 

\section{Bistability induced by time-delay}
\label{sec:bistability}

\begin{figure}[t]
\centering
\includegraphics[width=0.85\textwidth]{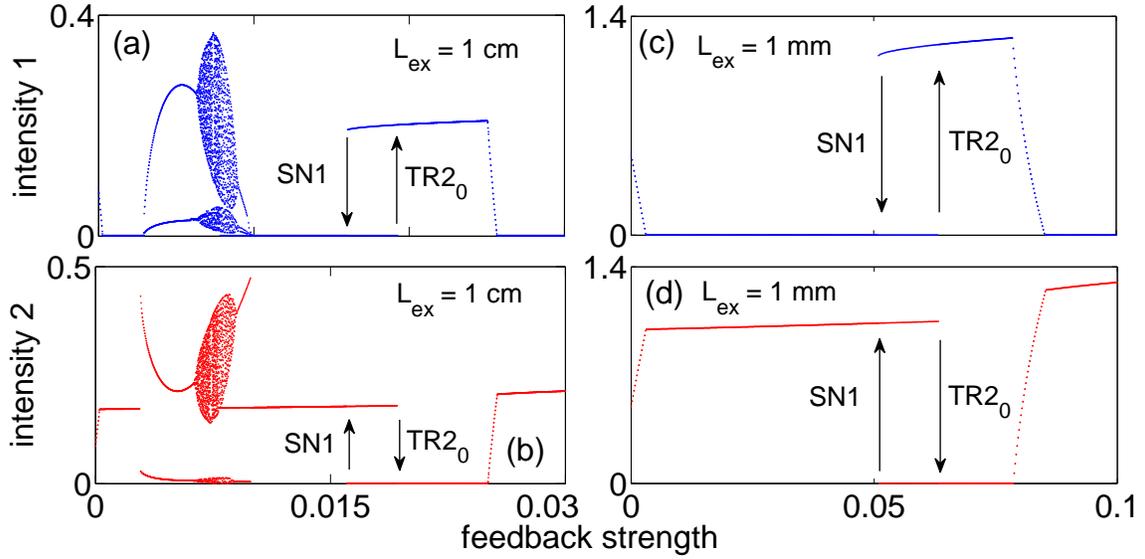}
\caption{(a), (b) Bifurcation diagrams of the modal intensities with varying feedback strength. Feedback 
parameters are as in Figure \ref{fig:equilibria_1cm}. The label SN1 indicates a saddle-node bifurcation 
for mode 1. The label $\mbox{TR2}_{\mbox{\tiny 0}}$ indicates a transcritical bifurcation between a single-mode 
ECM of mode 2 associated with the free running solution and a dual-mode equilibrium. (c), (d) Same as Figure 
\ref{fig:bifurcation_diagrams} (a) and \ref{fig:bifurcation_diagrams} (b) but with an external cavity length 
of 1 mm.}
\label{fig:bifurcation_diagrams}
\end{figure}

In the dual-mode system described by equations (\ref{eq:LK_rate_equations}) bistability between single-mode ECMs 
can be observed. This is shown in the bifurcation diagrams of the modal intensities in Figure 
\ref{fig:bifurcation_diagrams} (a) and \ref{fig:bifurcation_diagrams} (b). Here the 
feedback parameters are the same as in Figure \ref{fig:equilibria_1cm}. In these figures one can see that, as the 
feedback strength is increased from zero, the system evolves into a stable single-mode ECM of mode 2, followed 
by a Hopf bifurcation and a region of coupled complex dynamics. For a further increase of the feedback strength 
a single-mode ECM of mode 2 is the only stable state until a  pair of single-mode ECMs of mode 1 are generated 
through a saddle-node bifurcation, labelled SN1. At this saddle-node bifurcation a region of bistability appears 
where the unique single-mode ECM of mode 2 and the single-mode ECM of mode 1 associated with the stable branch 
of the saddle-node bifurcation SN1 co-exist. This bistable region is bounded when the single-mode ECM of mode 2 
loses stability at a point which we label $\mbox{TR2}_{\mbox{\small 0}}$. Following this, the single-mode ECM of 
mode 1 is the only stable state of the system. For a further increase of the feedback strength we found a 
stable dual-mode equilibrium followed by a stable single-mode ECM of mode 2. 

A bistable region between single-mode ECMs can also be found for a system with 1 mm external cavity length, 
as one can see in Figure \ref{fig:bifurcation_diagrams} (c) and \ref{fig:bifurcation_diagrams} (d). This is 
a limiting case appropriate to a dual-mode microlaser with
a cavity length of order one hundred micrometres \cite{Zhukovsky2009, Chen2011}.
In this case,  we have increased $\gamma$ by a factor of three to reflect the possibility of increased optical 
losses in the microlaser system. This leads to an increase of the current threshold density $J_{thr} = \gamma / 
(g \tau_s)$ by a factor of three. We also choose a pump current five times greater than the current threshold, 
which leads to an increase of the relaxation oscillation frequency $\omega_{ro} = \sqrt{g J - \gamma / \tau_s}$ 
and of the damping factor $\sigma_{ro} = - g J/(2 \gamma)$ of the free running device. The increased relaxation 
oscillation frequency and damping can lead to faster switching transients for the memory.

\begin{figure}[t]
\centering
\includegraphics[width=0.9\textwidth]{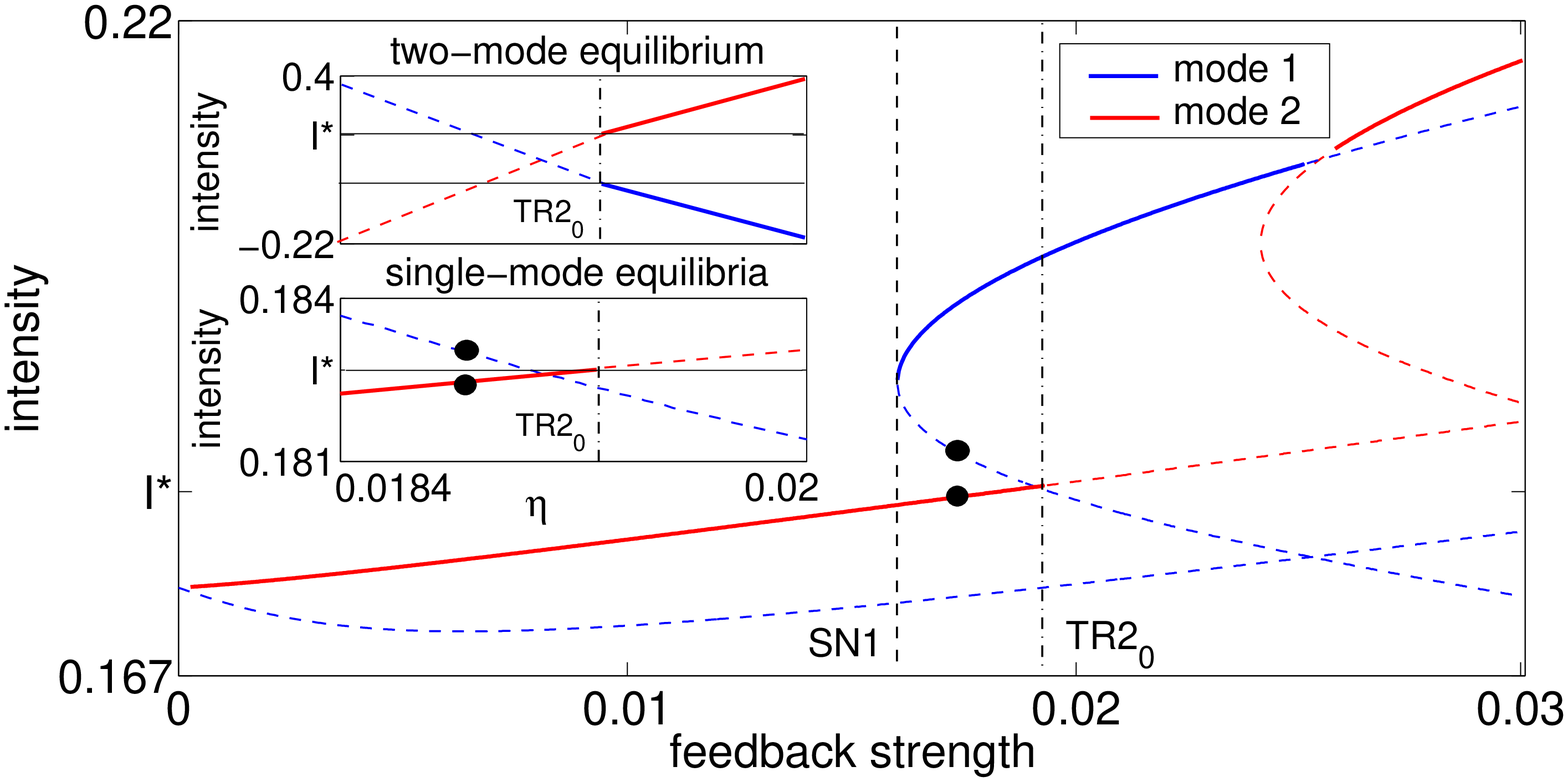}
\caption{Modal intensities of the single-mode ECMs with 1 cm external cavity length. Continuous and dashed 
lines indicate respectively stable and unstable equilibria. The labels SN1 and $\mbox{TR2}_{\mbox{\tiny 0}}$ have 
the same meaning of Figure \ref{fig:bifurcation_diagrams}. In the upper inset the modal intensities of the 
dual-mode solution involved in the bifurcation $\mbox{TR2}_{\mbox{\tiny 0}}$ are plotted.  This dual-mode solution 
is derived from the two single-mode ECMs indicated by the filled circles and plotted in the lower inset.  The 
label $I^*$ indicates the intensity of mode 2 at the bifurcation $\mbox{TR2}_{\mbox{\tiny 0}}$.}
\label{fig:steady_states_with_transverse_stability_1cm}
\end{figure}

In order to understand the bifurcation at the upper boundary of the bistable region, in Figure 
\ref{fig:steady_states_with_transverse_stability_1cm} we have plotted the intensity of the single-mode ECMs of 
interest for a device with 1 cm external cavity length. One can see the branches of single mode ECMs of mode 1, 
which appear at the saddle-node bifurcation SN1. In the upper inset of Figure 
\ref{fig:steady_states_with_transverse_stability_1cm} we plot the modal intensities of a dual-mode solution of 
the system in a range close to the bifurcation $\mbox{TR2}_{\mbox{\tiny 0}}$. As indicated by the filled circles 
in Figure \ref{fig:steady_states_with_transverse_stability_1cm}, this dual-mode solution is derived from the 
single-mode ECM of mode 1 associated with the unstable branch of the bifurcation SN1 and from the single-mode 
ECM of mode 2 associated with the free running solution. The modal intensities of these single-mode ECMs have 
been plotted in the lower inset of Figure \ref{fig:steady_states_with_transverse_stability_1cm}. From the two 
insets we can see that the upper boundary of the bistable region is found where one of the fields associated 
with the dual-mode solution becomes equal to zero. The upper boundary of the bistable region is then 
understood as resulting from a collision between a stable single-mode ECM of mode 2 with an unstable dual-mode 
solution. These solutions exchange their stability in a transcritical bifurcation $\mbox{TR2}_{\mbox{\tiny 0}}$ 
with the dual-mode solution becoming unphysical with a negative value for the intensity of mode 1 after the 
bifurcation.

Stability of single-mode ECMs in the larger dual-mode system requires that they be stable with respect to 
perturbations in the direction transverse to the single-mode sub-manifolds of the system. Here, stability requires 
that eigenvalues of the zero field part of the single-mode solutions are transversally stable with respect to 
the carrier density value given by equation (\ref{eq:single_mode_equilibrium_feedback}). 
As the amplitude of the perturbed field is zero, the problem is linear and leads to the following condition for 
transverse stability of single-mode ECMs:
\setlength{\arraycolsep}{0.0em}
\begin{eqnarray}
\label{eq:condition_transverse_stability}
&& \mbox{Re}(\tilde{s}_m) = \mbox{Re}(W( \tilde{B}_m \tau \exp(-\tilde{A}_m \tau) )/\tau + \tilde{A}_m) < 0.
\end{eqnarray}
\setlength{\arraycolsep}{5pt}
Here $\mbox{Re}(\tilde{s}_m)$ denotes the real part of the complex eigenvalue $\tilde{s}_m$,  $W(X)$ is the 
Lambert function defined as the solution of the equation $x = W(x) \exp W(x)$  \cite{Andreas2007}. 
$\tilde{A}_m$ is defined such that $\tilde{A}_1 = (1/2)(1+i\alpha)[g_2^0(2n^{\hbox{\tiny sm}}+1)/(1+\epsilon\beta I_2^{\hbox{\tiny sm}})-1]$ 
and  $\tilde{A}_2 = (1/2)(1+i\alpha)[g_1^0(2n^{\hbox{\tiny sm}}+1)/(1+\epsilon\beta I_1^{\hbox{\tiny sm}})-1]$. $\tilde{B}_m$ is defined 
such that $\tilde{B}_1 = \eta\exp(-i(\varphi+\Delta\varphi))$ and $\tilde{B}_2 = \eta\exp(-i\varphi)$.
Equation (\ref{eq:condition_transverse_stability}) determines an additional stability criterion, which together with 
equation (\ref{eq:saddle_node_condition}) determines necessary conditions for stability of the single-mode ECMs of 
the system. Continuous and dashed lines in Figure \ref{fig:steady_states_with_transverse_stability_1cm} respectively 
denote stable and unstable single-mode ECMs obtained in this way.

\section{Bistable regions and role of the gain saturation}
\label{sec:bistable_regions}

\begin{figure}[t]
\centering
\includegraphics[width=0.9\textwidth]{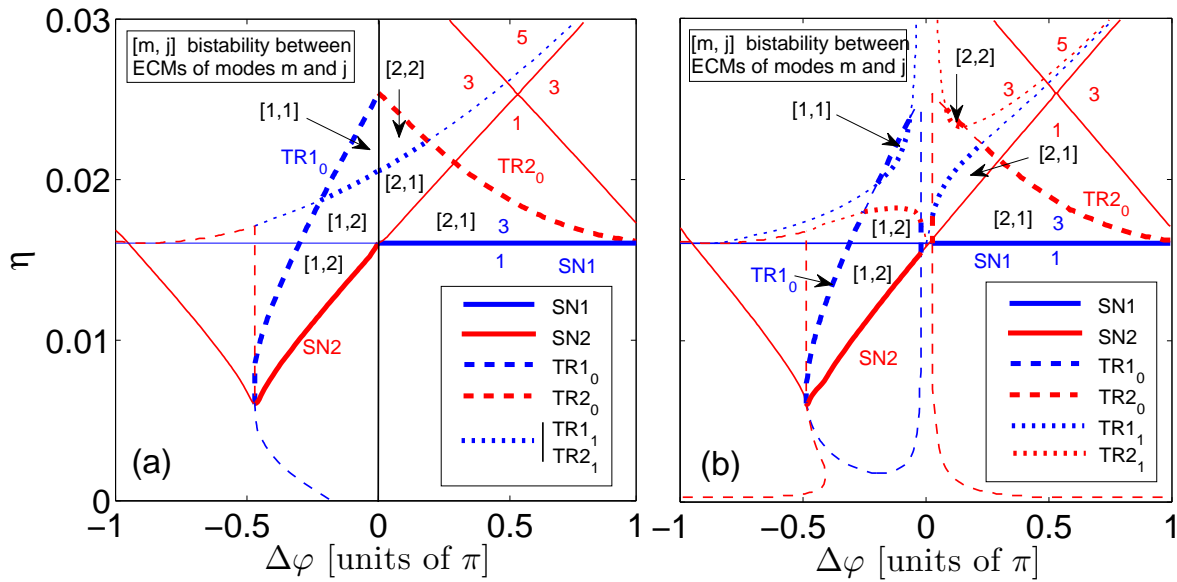}
\caption{(a) Map of the bistable regions as a function of feedback strength $\eta$ and the feedback phase 
difference $\Delta\varphi$ with $\epsilon=0$ and $\varphi=1.1\pi$. Feedback parameters are as in Figure 1. 
The bifurcation lines of single-mode ECMs of mode 1 and mode 2 that determine the structure of the bistable
region are plotted. Lines corresponding to the boundaries of the bistability region are plotted in bold. In 
each region, the notation $[m,j]$ indicates bistability between single-mode ECMs of mode m, and ECMs of mode 
j. The labels SNm and $\mbox{TRm}_{\mbox{\tiny j}}$ denote respectively saddle-node and transcritical bifurcations 
for mode m. (b) Same as Figure \ref{fig:bifurcation_map_delta_phi} (a), but with $\epsilon=0.01$.}
\label{fig:bifurcation_map_delta_phi}
\end{figure}

A complete understanding of the stability of the single-mode ECMs in the dual-mode system considered here can be 
obtained through equations (\ref{eq:saddle_node_condition}) and (\ref{eq:condition_transverse_stability}). We have 
used these equations to plot the boundaries of the bistable region varying the feedback strength and the feedback 
phase difference. The results are shown in Figure \ref{fig:bifurcation_map_delta_phi} (a) for a system with 1 cm 
external cavity length and $\epsilon = 0$. In figure \ref{fig:bifurcation_map_delta_phi} (a) only the bifurcation 
lines related to the boundaries of the bistable regions have been plotted. The bistable regions have been labeled 
using the notation $[m,j]$ which indicates bistability between single-mode ECMs of mode m and ECMs of mode j. The 
label $\mbox{SNm}$ indicates a saddle-node bifurcation of mode m. The label $\mbox{TRm}_{\mbox{\small 0}}$ denotes a 
transcritical bifurcation of mode m between a single-mode ECM associated with the free running solution and a 
dual-mode equilibrium. The label $\mbox{TRm}_{\mbox{\small 1}}$ denotes a transcritical bifurcation for mode m between 
a single-mode ECM born in the first saddle-node bifurcation of mode m and a dual-mode equilibrium. The integer 
numbers 1, 3, 5 indicate the number of single-mode ECMs of the system following each saddle-node bifurcation.

In Figure \ref{fig:bifurcation_map_delta_phi} (a) we observe four bistable regions. For lower values of $\eta$ 
there are two large bistable regions, labelled $[1,2]$ and $[2,1]$, which are characterized by a structure of 
the single-mode ECMs similar to the plots of Figure \ref{fig:steady_states_with_transverse_stability_1cm}. In 
this case, the label $[1,2]$ indicates that the ECMs generated by the saddle-node bifurcation are associated 
with mode 2, while the label $[2,1]$ indicates that these ECMs are associated with mode 1. For higher values 
of $\eta$ we found the bistable regions labelled $[1,1]$ and $[2,2]$, where we found bistability between ECMs 
of the same mode. These regions can be the basis of an optical memory where the wavelength channels are almost
degenerate in frequency.

A non-zero gain saturation parameter $\epsilon$ can influence the position of the bifurcations at the boundaries 
of the bistable regions. This can be understood by comparing Figure \ref{fig:bifurcation_map_delta_phi} (a) with 
Figure \ref{fig:bifurcation_map_delta_phi} (b), where the bifurcations at the boundaries of the regions have been 
plotted for the device with 1 cm external cavity length and $\epsilon = 0.01$. The main effect of the introduction 
of a non-zero gain saturation parameter is that the four bistable regions become non-adjacent. This separation of 
the bistable regions is due to the absence of the degeneracy between the bifurcations $\mbox{TR1}_{\mbox{\small 1}}$ 
and $\mbox{TR2}_{\mbox{\small 1}}$ and between $\mbox{TR1}_{\mbox{\small 0}}$ and $\mbox{TR2}_{\mbox{\small 0}}$. In order to 
understand the physical mechanism that governs the degeneracy of these bifurcations we study the dual-mode 
equilibrium of equations (\ref{eq:LK_rate_equations}) for $\epsilon = 0$, which can be written as follows:
\setlength{\arraycolsep}{0.0em}
\begin{eqnarray}
\label{eq:two_mode_equilibrium_feedback_zero_epsilon}
& n^{\hbox{\tiny tm}}_0 = - \eta \cos(\Delta\omega_1^{\hbox{\tiny sm}} \tau + \varphi_1) = - \eta 
\cos(\Delta\omega_2^{\hbox{\tiny sm}} \tau + \varphi_2) \nonumber\\
& p - n^{\hbox{\tiny tm}}_0 - (2 n^{\hbox{\tiny tm}}_0 + 1)( I_1^{\hbox{\tiny tm}} + I_2^{\hbox{\tiny tm}} ) = 0.
\end{eqnarray}
\setlength{\arraycolsep}{5pt}
Equations (\ref{eq:two_mode_equilibrium_feedback_zero_epsilon}) define a unique value of the excess carrier 
density and a unique value of the sum of the modal intensities, but do not uniquely define the two individual 
modal intensities. This means that these equations are solved by a continuous set of values of the modal 
intensities. Therefore equations (\ref{eq:two_mode_equilibrium_feedback_zero_epsilon}) define an invariant 
set of points in the phase space of the dual-mode system. We note also that the two conditions that define 
the dual-mode carrier density in equations (\ref{eq:two_mode_equilibrium_feedback_zero_epsilon}) coincide 
with the expressions for the single-mode carrier densities in equations (\ref{eq:single_mode_equilibrium_feedback}) 
for $\epsilon = 0$. Physically, this means that for $\epsilon = 0$ the dual-mode equilibrium states exist only 
when the excess carrier densities associated with the single-mode equilibria are degenerate. Thus the invariant 
set of dual-mode equilibrium states exist at isolated points where they enable the transcritical bifurcations 
at the boundaries of the bistable regions.

\begin{figure}[t]
\centering
\includegraphics[width=0.9\textwidth]{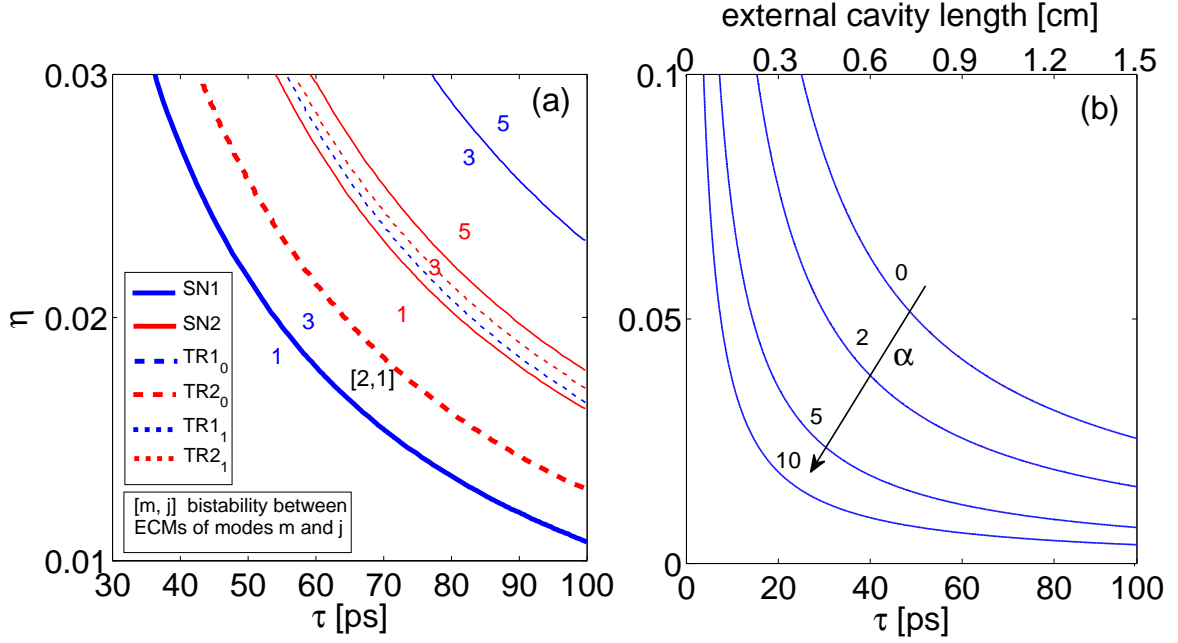}
\caption{(a) Map of the bistable region as a function of feedback strength $\eta$ and the delay time $\tau$ 
with $\epsilon=0.01$. Lines and labels have the same meaning of Figure \ref{fig:bifurcation_map_delta_phi}. 
(b) Saddle-node line for mode 1 as a function of feedback strength $\eta$ and the delay time $\tau$ for 
different values of $\alpha$ with $\epsilon=0.01$.}
\label{fig:bifurcation_map_tau}
\end{figure}

Equations (\ref{eq:saddle_node_condition}) and (\ref{eq:condition_transverse_stability}) also enable us to 
study the width of the bistable region as the delay time is decreased. In this case we have found that 
although the bistable region remains open for vanishing delay time, the feedback strength required to observe 
the bistability diverges in this limit. This is shown in Figure \ref{fig:bifurcation_map_tau} (a) where we 
plot the transcritical and saddle-node bifurcations of the single-mode ECMs for varying external cavity 
length and other parameters as in Fig. \ref{fig:steady_states_with_transverse_stability_1cm}. In this figure 
there is only one bistable region with a structure similar to the bistability presented in Figure 
\ref{fig:steady_states_with_transverse_stability_1cm}. The saddle-node line of mode 1 that opens the bistable 
region plotted in Figure \ref{fig:bifurcation_map_tau} (a) defines the minimum feedback strength and the 
minimum delay time required for observing this bistablility. It is interesting to note also that these 
minimum values can be reduced increasing the $\alpha$ parameter. This is shown in Figure 
\ref{fig:bifurcation_map_tau} (b) where the lower boundary of the bistable region has been plotted for 
different values of  $\alpha$.

\section{Switching dynamics of the optical memory}
\label{sec:switching_dynamics}

\begin{figure}[t]
\centering
\includegraphics[width=0.9\textwidth]{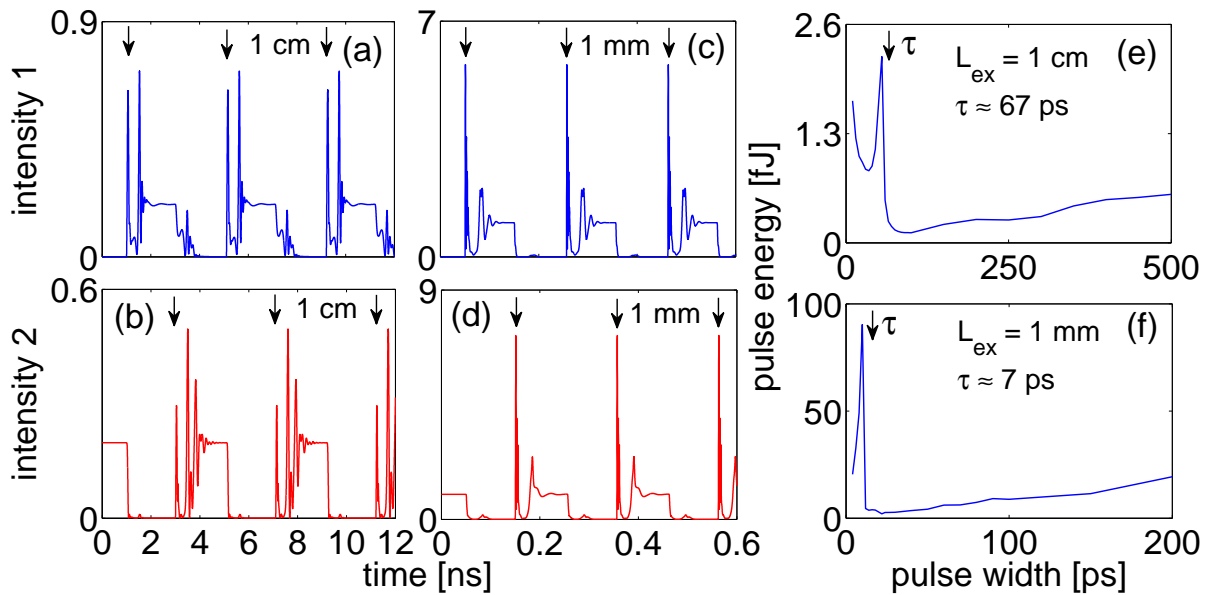}
\caption{(a), (b) Switching dynamics of the modal intensities with injected optical pulses at the instants 
indicated by the arrows. The memory switching frequency is 500 MHz. Feedback parameters are as in Figure 1. 
(c), (d) Switching dynamics of the modal intensities with a 1 mm external cavity length. The memory 
switching frequency is 5 GHz.  (e) Minimum energy of the injected pulse required for switching the memory 
as function of the pulse duration with a 1 cm external cavity length. The energy scale has been defined 
with respect to 1 mW output power of the system. The vertical arrow indicates the delay time of the system. 
(f) Same as Figure \ref{fig:memory_switching} (e), but with a 1 mm external cavity length}
\label{fig:memory_switching}
\end{figure}

The dual-mode system described by the equations (\ref{eq:LK_rate_equations}) in the bistable regime can be 
used as optical memory when optical pulses are injected in this system at frequencies corresponding to the 
detunings of the external cavity modes. In Figure \ref{fig:memory_switching} (a) and \ref{fig:memory_switching} 
(b) we have studied the switching dynamics of a system with external cavity length of 1 cm. We have used the 
parameters $\Delta\omega_1^{\mbox{\tiny inj}} = -7.39$ GHz, $\Delta\omega_2^{\mbox{\tiny inj}} = 0.38$ GHz,  
$\eta = 0.017$,  $k_1 = k_2 = 0.022$. Optical pulses with a duration of 100 ps have been injected at the 
instants indicated by the arrows. The memory is switched with a frequency of 500 MHz with a pump current at 
35 $\%$ above threshold. The two modes show transients with the presence of several fast spikes. In Figure 
\ref{fig:memory_switching} (c) and \ref{fig:memory_switching} (d) the switching dynamics of the modal 
intensities have been plotted with an external cavity length of 1 mm. The parameters used are 
$\Delta\omega_1^{\mbox{\tiny inj}} = -74.28$ GHz, $\Delta\omega_2^{\mbox{\tiny inj}} = 3.80$ GHz, $\eta = 0.056$, 
$k_1 = k_2 = 0.054$ and $J/J_{thr} = 5$. The memory is switched with a frequency of 5 GHz using optical 
pulses with a duration of 10 ps. This increase of one order of magnitude of the switching speed is 
consistent with the increase of the relaxation oscillation frequency due to the increase of the pump 
current and of the cavity decay rate.

In Figure \ref{fig:memory_switching} we plot the minimum pulse energy as a function of the pulse duration 
for systems with external cavity lengths of 1 cm and 1 mm. We have defined this minimum pulse energy by 
requiring that the switching transients be strongly damped before the arrival of the next switching pulse. 
For pulse durations larger than the delay time, we observe that the energy required for switching the system 
decreases linearly as the pulse duration is decreased. The minimum pulse energy in each case corresponds to 
a pulse duration close to but greater than the delay time. This result can be understood  when we consider 
that the switching pulse is injected in a mode of the system which has zero field amplitude. If the pulse 
duration is longer than the external cavity round trip time then the leading edge of the pulse interacts 
with the system and the output is fed back before the arrival of the trailing edge. The zero field mode 
is then continuously fed with energy during the switching process. In Figure \ref{fig:memory_switching} 
(e) and \ref{fig:memory_switching} (f) the optimal pulse energy we obtain is of a few femtojoules in each 
case. These results suggest that feedback induced bistability in microlaser systems can lead to compact 
all-optical memories with switching speeds in the GHz range and injected pulse energies of a few femtojoules.

\section{Conclusions}
In conclusion, we have presented an analytical and numerical study of the bifurcation structures of the 
bistability between single-mode equilibria in dual-mode diode laser systems with time delayed optical 
feedback.  We have proposed an all-optical memory based on this bistability, which shows attractive 
features in the context of in-plane all-optical signal processing, including the short external cavity 
length, the self-sustained nature of the feedback system, and the symmetry of the memory states. The 
optimal pulse duration has been shown to be determined by the delay time of the system. With feedback 
parameters appropriate to dual-mode devices based on conventional edge-emitting diode lasers, switching 
frequencies of at least 500 MHz with 100 ps pulses should be possible. This would represent a significant 
improvement on the performance of a memory element based on the injection locking bistability in a
conventional device.  We also considered the limit of a very short external cavity length equal to 1 mm, 
which may be appropriate for certain photonic crystal microlaser structures. In this case we obtained 
memory switching frequencies up to 5 GHz using 10 ps injected optical pulses and with pulse energies of 
a few femtojoules.

\section*{Acknowledgements}
The authors wish to thank Science Foundation Ireland for financial support of this work.


\bibliographystyle{IEEEtran}

\end{document}